\documentstyle[12pt]{article}

\begin{document}

{\bf Mass number dependence in }$\Lambda -${\bf hypernuclei by means of the
Hypervirial Theorems}

\begin{center}
Th.E.Liolios\footnote{%
email:theoliol@ccf.auth.gr}

{\footnotesize Department of Theoretical Physics,University of
Thessaloniki,Thessaloniki 54006,Greece}
\end{center}

{\bf Abstract}

The Hypervirial Theorems are applied to a wide class of single particle
nuclear potentials, in order to study the mass number dependence of various
quantities in $\Lambda -$hypernuclei. A very efficient model is assumed for
the radius parameter of the potentials and the limitations of the whole
method are discussed.

\vspace{0.5cm} {\it PACS: 21.80.+a, 13.75.Ev, 21.10.Dr}\vspace{0.5cm}

{\bf I. Introduction}

Over the past years a variety of methods has been used for the determination
of the $\Lambda -$hyperon binding energies in hypernuclei, ranging from that
of nuclear emulsions to the strangeness exchange reaction\cite
{povh,bruckner,bertini1} up to the more recent associated production reaction%
\cite{milner,chrien1,pile1,chrien2,pile2}. Those experiments have yielded a
multitude of experimental results which allow for a better description of
the mass number dependence of the $\Lambda -$binding energies. Moreover,
this wealth of data provides a solid foundation of an analytical approach of 
$\Lambda -$hypernuclei. Various single particle potentials have been used in
the past in order to approximate the self-cosistent potenial in a $\Lambda $%
-hypenucleus \cite{millener,lal93,lal94,lalg,lalomega,grypprob}, while their
parameters have been studied systematically by means of the associated
production reaction $\left( \pi ^{+},K^{-}\right) .$ The results indicate
that the depth of the potential can be safely regarded as mass number
independent (except for very light nuclei) with a value close to $30MeV$\cite
{grypprob,dalitz,bertini2,bou,gal}$.$ However, it turns out that mass number
dependence has to be taken seriously into account for a nuclear potential to
reproduce the experimental data satisfactorily. With data available on the
spacing of single particle levels in a given hypernuclei, one can also
determine the functional dependence of the radius parameter on the mass
number, thus constructing a very efficient potential model.

In addition, in a recent work, the Hypervirial Theorems (HVT) were applied
to a wide class of single patricle nuclear potentials yielding various
valuable approximate quantities \cite{jag,lioint,cpc,gryplio}. Hence, by
incorporating a mass number dependent radius parameter in that work a
thorough study of $\Lambda -$hypernuclei can be accomplished.

{\bf II. Application of the hypervirial theorems to single particle nuclear
potentials.}

A wide class of potentials used in Nuclear Physics is : 
\begin{equation}
\label{vf}V(r)=-Df(\frac rR)\hspace{1cm}0\leq r<\infty 
\end{equation}
where $D$ is the potential depth, $R=r_0A^{\frac 13}$ the potential radius, $%
r_0$ the radius parameter and $A$ the mass number of the nucleus. The
''potential form factor'' f, ($f(0)=1),$ is an even analytic function of $%
x=\frac rR.$ Namely, 
\begin{equation}
\label{fdx}f(x)=\sum\limits_{k=0}^\infty d_kx^{2k} 
\end{equation}
where $d_k$ are the numbers: 
\begin{equation}
\label{dk}d_k=\frac 1{(2k)!}\frac{d^{2k}}{dx^{2k}}f(x)\mid _{x=0}\hspace{1cm}%
k=0,1,2,...\hspace{1cm}d_1<0 
\end{equation}

The corresponding Schroedinger eigenvalue problem is:

$$
\left[ \frac{d^2}{dx^2}-\frac{l(l+1)}{x^2}+s^{-2}f\left( x\right) +\stackrel{%
\sim }{E}_{nl}\right] u_{nl}\left( x\right) =0 
$$
\begin{equation}
\label{schs}u_{nl}\left( 0\right) =0\qquad u_{nl}\left( \infty \right) =0 
\end{equation}

where, $x=\frac rR,$ $s=\left( \frac{\hbar ^2}{2\mu DR^2}\right) ^{\frac 12}$
and $\stackrel{\sim }{E}_{nl}=s^{-2}\frac{E_{nl}}D$

An application of the hypervirial theorems to$\left( \ref{schs}\right) $
yielded fairly accurate series for the corresponding energy eigenvalues $%
E_{nl}$, the potential $<V>_{nl}$ and the kinetic energy $<T>_{nl}$ and the
mean-square radius (MSR) $<r^2>_{nl}$. By observing\cite{gryplio} that those
quantities are actually exclusive function of the parameter $s$ one can
write those series in the form: 
\begin{equation}
\label{hvten}E_{nl}=D\sum_{k=0}^\infty e_{nl}^{\left( k\right) }s^k\qquad
,\qquad <V>_{nl}=D\sum_{k=0}^\infty v_{nl}^{\left( k\right) }s^k\qquad
,\qquad <T>_{nl}=D\sum_{k=0}^{\infty t}t_{nl}^{\left( k\right) }s^k 
\end{equation}

\begin{equation}
\label{hvtrad}<r^2>_{nl}=R^2\sum_{k=0}^\infty r_{nl}^{\left( k\right) }s^k 
\end{equation}

where

\begin{equation}
\label{tv}t_{nl}^{\left( k\right) }=\left( \frac{1+k}2\right) e_{nl}^{\left(
k\right) }\qquad ,\qquad v_{nl}^{\left( k\right) }=\left( \frac{1-k}2\right)
e_{nl}^{\left( k\right) } 
\end{equation}

The first significant coefficients $e_{nl}^{\left( k\right) }$ are :

$$
e_{nl}^{\left( 0\right) }=-1\qquad ,\qquad e_{nl}^{\left( 1\right)
}=2a_{nl}\left( -d_1\right) ^{\frac 12}\qquad ,\qquad e_{nl}^{\left(
2\right) }=\frac{d_2}{8d_1}\left[ 12a_{nl}^2-4l(l+1)+3\right] 
$$

\begin{equation}
\label{ek}e_{nl}^{\left( 3\right) }=-\frac{a_{nl}(-d_1)^{\frac 12}}{32d_1^3}%
\left\{ 4a_{nl}^2\left( 20d_1d_3-17d_2^2\right) +4d_1d_3\left[
25-12l(l+1)\right] +d_2^2\left[ 36l(l+1)-67\right] \right\} 
\end{equation}

Likewise, the corresponding first significant coefficients for the MSR are:

$$
r_{nl}^{\left( 0\right) }=0\qquad ,\qquad r_{nl}^{\left( 1\right) }=\frac{%
a_{nl}}{(-d_1)^{\frac 12}}\qquad ,\qquad r_{nl}^{\left( 2\right) }=\frac{d_2%
}{8d_1^2}\left[ 12a_{nl}^2-4l(l+1)+3\right] 
$$
\begin{equation}
\label{rk}r_{nl}^{\left( 3\right) }=-\frac{a_{nl}(-d_1)^{\frac 12}}{64d_1^4}%
\left\{ 20a_{nl}^2\left( 12d_1d_3-17d_2^2\right) +12d_1d_3\left[
25-12l(l+1)\right] +5d_2^2\left[ 36l(l+1)-67\right] \right\} 
\end{equation}

{\bf III. Derivation of a general formula for mass number dependent
quantities.}

Potentials of the form $\left( \ref{vf}\right) $ have been used extensively
in Hypernuclear Physics with very satisfactory results. Nevertheless , mass
number dependence has to be assumed for the radius parameter $r_0$ as well,
for a reliable reproduction of the experimental results, that is:

\begin{equation}
\label{r0}R=r_0\left( A\right) A^{\frac 13} 
\end{equation}

where $A$ the mass number of the core nucleus , assuming the rigid core
model. Such an assumption effectively implies that the volume each nucleon
occupies $\left( \frac 43\pi r_0^3\right) $ as well as the mean
interparticle distance $\left( 2r_0\right) $ are functions of $A$ , too.
According to some very thorough studies in the field of hypernuclei the
experimental results can be fitted much better by a single particle
potential if the radius parameter $r_0$ is assumed to have the following form%
\cite{millener,lal94}:

\begin{equation}
\label{r01}r_0=\left( r_1+r_2A^{-\frac 23}\right) 
\end{equation}

On the basis of this assumption a member of the potential class$\left( \ref
{vf}\right) $ was employed to study the functional dependence of binding
energies, kinetic and potential energies in $\Lambda -$hypernuclei\cite
{lal94}. That study can now be generalised, by means of the HVT, for the
entire class$\left( \ref{vf}\right) $, including the study of the MSR.

The parameter $s$ appearing in the previous section can now be written :

\begin{equation}
s=s_\Lambda \frac{\sqrt{1+gA^{-1}}}{A^{\frac 13}\left( 1+bA^{-\frac
23}\right) }, 
\end{equation}

where the new parameters introduced are:

\begin{equation}
s_\Lambda =\sqrt{\frac{\hbar ^2}{2m_\Lambda Dr_0^2}},\qquad b=\frac{r_2}{r_1}%
,\qquad g=\frac{m_\Lambda }{m_N} 
\end{equation}

$m_N,m_\Lambda $ are the masses of the nucleon and of the $\Lambda -$%
particle respectively.

By expanding $s$ in terms of $A^{-\frac 13}$ and dropping terms of order
higher than $A^{-1}$ the following approximation is arrived at:

\begin{equation}
s\simeq s_\Lambda \left( A^{-\frac 13}-bA^{-1}\right) 
\end{equation}

Consequently the expectation value of a quantity $Q,$ using the HVT
formalism, takes the approximate form:

\begin{equation}
\label{qsum}<Q>_{nl}=\sum_{k=0}^\infty q_{nl}^{\left( k\right) }s^k\simeq
\sum_{k=0}^\infty q_{nl}^{\left( k\right) }s_\Lambda ^k\left( A^{-\frac
13}-bA^{-1}\right) ^k 
\end{equation}

Neglecting higher terms , the above approximation is finally written as
follows:

\begin{equation}
\label{qexp}<Q>_{nl}\simeq q_{nl}^{\left( 0\right) }+2s_0q_{nl}^{\left(
1\right) }A^{-\frac 13}+4s_0^2q_{nl}^{\left( 3\right) }A^{-\frac 23}+\left(
8s_0^3q_{nl}^{\left( 3\right) }-2bs_0q_{nl}^{\left( 1\right) }\right) A^{-1} 
\end{equation}
where $s_0=\frac{s_\Lambda }2$

By replacing $q_{nl}^{\left( k\right) }$ in the above expression with the
corresponding coefficients $e_{nl}^{\left( k\right) }$, $v_{nl}^{\left(
k\right) },t_{nl}^{\left( k\right) }$, $r_{nl}^{\left( k\right) },$ the mass
number dependence of $E_{nl},<V>_{nl},<T>_{nl},$ and $<r^2>_{nl}$ assumes a
very enlightening quantitative form for each state.

As $g$ vanishes in the present approximation , for $b=0$ formula $\left( \ref
{qexp}\right) $ corresponds to the conventional model introduced by the
nuclear radius $R=r_0A^{\frac 13}.$

At this venture some comments regarding the MSR are deemed necessary as the
importance of the $A$ dependence of the energy quantities is obvious,
anyway. Although there are no experimental data for the MSR of a $\Lambda -$%
particle,its usefullness becomes apparent in the calculation of the
oscillator level spacing according to the approximate formula:

\begin{equation}
\label{hbaro}\hbar \omega _\Lambda =\frac 32\frac{\hbar ^2}\mu
<r^2>_{00}^{-1} 
\end{equation}

A thorough study of the validity of the above formula was undertaken\cite
{lalomega} using the reduced Poeschl-Teller potential which behaves very
much alike the Gaussian one , with some extra analytic features.

Moreover, the oscillator spacing can be used in the calculation of the
relative probability of the recoiless $\Lambda -$production using the
Debye-Waller factor\cite{grypprob}:

\begin{equation}
\label{pnn}P\left( n_i,n_i\right) \simeq \exp \left[ \frac{-\left(
2n_i+1\right) \hbar ^2q^2}{2m_\Lambda \hbar \omega _\Lambda }\right] 
\end{equation}

where $n_i$ is the quantum number of the harmonic oscillator state and $%
\hbar q$ the $\Lambda -$recoil momentum.

For the potential model used in \cite{lalomega,grypprob} the calculation of $%
\hbar \omega _\Lambda $ as a function of $A$ was easy and performed in a
straightforward way. In general, however, one has to resort to a special
treatment of potentials $\left( \ref{vf}\right) ,$ such as that of the HVT
which yields Eq.$\left( \ref{hvtrad}\right) .$

Hence, an assessment of Eq.$\left( \ref{hvtrad}\right) $ with respect to
existing experimental data can only be accomplished indirectly , through the
oscillator spacing and only for the ground state. Alternatively, a numerical
integration of the Schroedinger equation$\left( \ref{schs}\right) $ can also
provide a measure of its accuarcy\cite{jag,lioint,cpc}. The orbital radius
of a $\Lambda -$particle, taken as the root mean square radius RMSR =$\sqrt{%
<r>_{nl}^2},$ can now be written as a fraction of the potential radius:

\begin{equation}
\label{rf}\sqrt{<r>_{nl}^2}=F\left( A\right) R\qquad ,\qquad F\left(
A\right) =\sqrt{\sum_{k=0}^\infty r_{nl}^{\left( k\right) }s^k}<1 
\end{equation}
{\bf IV. Implementation of of the derived formulae and their limitations}

Two typical members of $\left( \ref{vf}\right) $ are the Gaussian potential :

\begin{equation}
\label{gauss}V\left( r\right) =-De^{-\frac{r^2}{R^2}} 
\end{equation}

for which

\begin{equation}
\label{gaud}d_k=\frac{\left( -1\right) ^k}{k!} 
\end{equation}

and the (reduced) Poeschl-Teller potential

\begin{equation}
\label{pteller}V\left( r\right) =-D\cosh ^{-2}\left( r/R\right) . 
\end{equation}

As for the Gaussian potential the previous formalism is readily applied to
it to obtain the mass number dependence of the binding energies :

\begin{equation}
\label{gaens}E_{1s}=-D\left[ 1-6s_0A^{-\frac 13}+\frac{15}2s_0^2A^{-\frac
23}+\left( \frac{25}8s_0^3+6bs_0\right) A^{-1}\right] 
\end{equation}

\begin{equation}
\label{gaenp}E_{1p}=-D\left[ 1-10s_0A^{-\frac 13}+\frac{35}2s_0^2A^{-\frac
23}+\left( \frac{105}8s_0^3+10bs_0\right) A^{-1}\right] 
\end{equation}

\begin{equation}
\label{gaend}E_{1d}=-D\left[ 1-14s_0A^{-\frac 13}+\frac{63}2s_0^2A^{-\frac
23}+\left( \frac{273}8s_0^3+14bs_0\right) A^{-1}\right] 
\end{equation}

\begin{equation}
\label{gaenss}E_{2s}=-D\left[ 1-14s_0A^{-\frac 13}+\frac{75}2s_0^2A^{-\frac
23}+\left( \frac{315}8s_0^3+14bs_0\right) A^{-1}\right] 
\end{equation}

Similarly one obtains the potential and kinetic energy mass number
dependence:

\begin{equation}
\label{gapots}V_{1s}=-D\left[ 1-3s_0A^{-\frac 13}-\left( \frac{25}{16}%
s_0^3-3bs_0\right) A^{-1}\right] 
\end{equation}

\begin{equation}
\label{gapotp}V_{1p}=-D\left[ 1-5s_0A^{-\frac 13}-\left( \frac{105}{16}%
s_0^3-5bs_0\right) A^{-1}\right] 
\end{equation}

\begin{equation}
\label{gakins}T_{1s}=D\left[ 3s_0A^{-\frac 13}-\frac{15}2s_0A^{-\frac
23}-\left( \frac{75}{16}s_0^3+3bs_0\right) A^{-1}\right] 
\end{equation}

\begin{equation}
\label{gakinp}T_{1p}=D\left[ 5s_0A^{-\frac 13}-\frac{35}2s_0A^{-\frac
23}-\left( \frac{375}{16}s_0^3+5bs_0\right) A^{-1}\right] 
\end{equation}

The above relations are identical to the ones obtained in a thorough study%
\cite{lal94} of the applicability of $\left( \ref{gauss}\right) $ to $%
\Lambda -$hypernuclei assuming the mass number dependence for the radius
parameter given by $\left( \ref{r01}\right) .$ In that work the validity of $%
\left( \ref{r01}\right) $ was firmly established by the substantial
agreement between the numerical values of the analytic expressions and the
corresponding experimental data.

The present approach can also provide the MSR for various states:

\begin{equation}
\label{gamsrs}<r^2>_{1s}=R^2\left[ 3s_0A^{-\frac 13}+\frac{405}{32}%
s_0^2A^{-\frac 23}+\frac{3\left( 135s_0^3-16bs_0\right) }{16}A^{-1}\right] 
\end{equation}

\begin{equation}
\label{gamsrp}<r^2>_{1p}=R^2\left[ 5s_0A^{-\frac 13}+\frac{1365}{32}%
s_0^2A^{-\frac 23}+\frac{5\left( 273s_0^3-16bs_0\right) }{16}A^{-1}\right] 
\end{equation}
Similar epressions can be produced for potential $\left( \ref{pteller}%
\right) $ , by means of $\left( \ref{qexp}\right) $.

In Fig.1, Fig.2 and Fig.3 , the binding energies of a $\Lambda -$hyperon and
the corresponding potential and kinetic energies are plotted with respect to 
$A^{-\frac 23}$. Available experimental data\cite{pile2} are also shown in
Fig.1. The fitting parameters for the Gaussian potential obtained are$\cite
{lal94}:$ $D=32.45MeV,r_1=1.438,r_2=-1.118$

In Fig$.4,$ Eqs. $\left( \ref{r0}\right) ,\left( \ref{gamsrs}\right) $, $%
\left( \ref{gamsrp}\right) $ (and the corresponding $<r^2>_{1d})$ are used
in order to plot the potential radius $R$ and the orbital radius of a $%
\Lambda -$hyperon, for various eigenstates, against $A^{\frac 13}.$ That
figure is in accordance with the argument that ''the distinguishable $%
\Lambda -$particle provides one of the best examples of single particle
shell structure in nuclear physics''\cite{millener}.

The method of the HVT introduces limitations to the present formalism from
the very beginning. A solid criterion for the practicality of the method is
the magnitude of the parameter $s.$ Apparently the truncated series $\left( 
\ref{qexp}\right) $ is accurate enough only if $s$ is sufficiently small
which is akin to the potential being deep and wide. In addition , previous
studies have shown that the method works better for the deeply bound states
of relatively heavy nuclei\cite{jag,lioint,cpc}. Therefore in Fig.4 the
reliable region is that lying lower than the potential radius as the more
the discontinuous curves approach the solid line from the right , the more
loosely bound the particle becomes.

An equally important limitation is due to the potentials $\left( \ref{vf}%
\right) $ being two-parameter ones and therefore unable to take into account
some surface effects a three-parameter potential such as the Woods-Saxon one
would. That inherent inadequacy renders such potentials unsuitable for the
description of very heavy nuclei. On the other hand, for vey light systems
the depth parameter should have also been assumed mass number dependent.

Consequently, for Eqs. $\left( \ref{gaens}\right) $ to $\left( \ref{gamsrp}%
\right) $, a safe validity range of hypernuclei is :$20<A<120,$ whereas, for
the loosely bound states, their accuracy could be slightly improved by
including a higher term (e.g. of order $A^{-\frac 43})$.

Finally, it should be underlined that nothing prevents the present formalism
from being applied to nucleon-nucleus interactions, possibly assuming a
different model for the radius parameter according to experimental data.

\begin{center}
{\bf ACKNOWLEDGEMENTS}
\end{center}

The author would like to thank C.Daskalogiannis, S.Massen and G.Lalazissis
for useful comments and discussion.

\end{document}